\documentstyle[twocolumn,prl,aps]{revtex}
\begin{document}
\twocolumn[\hsize\textwidth\columnwidth\hsize\csname @twocolumnfalse\endcsname

\draft

\title{
Generation of Large Moments in a Spin-1 Chain
with Random Antiferromagnetic Couplings}

\author{Kun Yang$^{1,2}$ and R. N. Bhatt$^1$} 
\address{$^1$Department of Electrical Engineering, Princeton University,
Princeton, NJ 08544}
\address{$^2$
Condensed Matter Physics 114-36, California Institute of Technology,
Pasadena, CA 91125}
\date{\today}
\maketitle

\begin{abstract}
We study the spin-1 chain with nearest neighbor couplings that
are rotationally invariant, but include both Heisenberg and
biquadratic exchange, with random strengths. We demonstrate, using
perturbative renormalization group methods as well as exact
diagonalization of clusters, that the system generates ferromagnetic
couplings under certain circumstances even when all the bare couplings
are antiferromagnetic.
This
disorder induced instability leads to formation of large magnetic moments
at low temperatures, and is a purely quantum mechanical effect
that
does not have a classical counterpart. The physical origin of this instability,
as well as its consequences, are discussed. 
\end{abstract}
\pacs{Pacs: 75.10.J, 75.30.H, 75.50.E}
]

Classical spin models have been studied extensively over the past half
century. The inclusion of quantum mechanical nature of the spin variables
result usually in {\em quantitative} effects such as 
renormalization of transition
temperatures, order parameter in the ordered phase, spin wave velocity, etc.
However, it began to be realized about two decades back that quantum
fluctuations could result in {\em qualitative} changes such as the nature
of the phase itself. Ma {\em et al.}\cite{dm} showed that in a spin-1/2
chain with nearest neighbor antiferromagnetic Heisenberg interactions of random
strength, quantum fluctuations lead to a divergent density of
excitations at low energy scales even in the absence of such divergence
in the distribution of bare couplings. Around the same time, Bhatt and
Lee \cite{bl} independently showed that this occurs for highly disordered
spin-1/2 antiferromagnets with short range interactions in higher dimensions as well, and that this phenomenon is
a purely quantum mechanical effect \cite{bh}. A few years later,
Haldane\cite{haldane} showed that the uniform integer spin chains with
antiferromagnetic Heisenberg interactions exhibited a gap, and concomitant
short-range spin-spin correlations in the ground state, in contrast with the
half integer as well as classical spin chains. The Haldane gap, which scales as
$\exp (-\pi S)$ with the spin $S$,
is also a purely quantum mechanical effect, unobtainable from
the classical ($S\rightarrow\infty$) limit, where it appears as an essential
singularity.
In the past few years, studies of the quantum Ising model in a transverse field
in one, two and three dimensions \cite{fisher2,rygbh} as well as Heisenberg
spin chains\cite{hyman1,hyman2,monthus} with randomness,
have demonstrated that the ground state ($T=0$) phase diagram 
includes a Griffiths
phase with 
divergent response functions due to rare fluctuations. Thus, electronic
systems in the quantum regime display a 
richer variety of phenomena resulting from the
interplay between correlation and disorder, than is usual for classical 
systems.

In this paper, we describe another phenomenon occurring in
spin-chains as a purely quantum mechanical effect - namely, the generation
of {\em ferromagnetic} (F) couplings, 
and consequently large moments leading to a Curie
susceptibility at low temperatures, in a spin-1 chain with isotropic
but random {\em antiferromagnetic} (AF) couplings. 
Though the generated moments become
arbitrarily large in the low temperature limit, the phenomenon of the
generation of ferromagnetic couplings relies on a purely quantum mechanical
effect, and has no classical analog.

For the random spin-1/2 chain with near neighbor 
AF interactions,
Fisher\cite{fisher1} showed that the real space renormalization group (RG)
scheme\cite{dm,bl} 
becomes asymptotically exact, and leads to a ``random singlet" phase, 
where distant pairs of spins form singlets in a hierarchical
manner dependent on the realization of the random bonds, and
dominate the low energy physics. In the presence of randomly placed
ferromagnetic couplings of arbitrary concentration, however,
Westerberg {\em et al.}\cite{westerberg} showed that the random
singlet phase is destroyed in one-dimension due to the formation of
large moments by active, ferromagnetically coupled spins, and the
magnetic susceptibility at
asymptotically low temperature assumes a pure Curie ($1/T$) form, right
upto the ferromagnetic point at zero concentration of antiferromagnetic bonds.

For a spin-1/2 system, spin rotational symmetry uniquely constrains the
coupling between spins to be the Heisenberg form\cite{note}, 
$J{\bf S}_i\cdot{\bf S}_j$. For spins with $S>1/2$, however,
the most general form of isotropic coupling between spins $i$ and $j$
with spin rotational invariance may be written in powers of the Heisenberg form:
$H_{ij}=\sum_{n=1}^{2S}J^{(n)}({\bf S}_i\cdot {\bf S}_j)^n.$
For the random spin-1 chain with nearest neighbor couplings, this implies
a Hamiltonian written most generally as:
\begin{eqnarray}
&H&=\sum_i[J_i{\bf S}_i\cdot {\bf S}_{i+1}+D_i({\bf S}_i
\cdot {\bf S}_{i+1})^2]\nonumber\\
&=&\sum_i\sqrt{J_i^2+D_i^2}[\cos\theta_i{\bf S}_i\cdot {\bf S}_{i+1}
+\sin\theta_i({\bf S}_i\cdot {\bf S}_{i+1})^2],
\end{eqnarray}
where the $J$'s and $D$'s are uncorrelated random variables.

With purely Heisenberg couplings, 
it was 
showed\cite{hyman2} that as long as there are no F bonds in the bare
Hamiltonian, the system cannot be in the pure Curie paramagnetic phase.
In this paper we show that this is {\em no longer} the case
when biquadratic couplings are present\cite{boechat}.

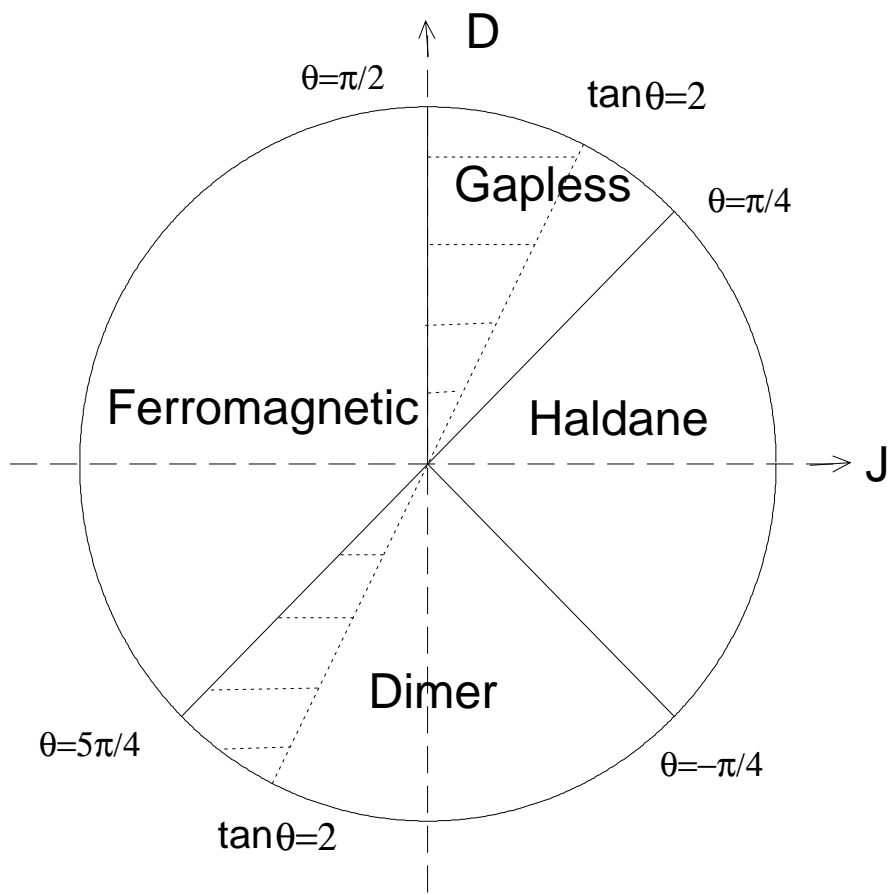
\begin{figure}
\caption{Phase diagram of a pure spin-1 chain. 
Solid lines are phase boundaries.
Shaded regions represent couplings in the antiferromagnetic 
sector satisfying $J-D/2 < 0$, which, in the presence of randomness,
could generate effective ferromagnetic bonds
at low energies.
}
\label{phase}
\end{figure}

With no randomness ({\em i.e.} $J_i = J, D_i = D$), the properties of the
Hamiltonian of Eq. (2) is controlled by the
angular variable $\theta$ satisfying $\tan\theta=D/J$. 
There exist four different 
phases (see Fig. \ref{phase}).
For $\pi/2 < \theta < 5\pi/4$, each
individual bond favors a total spin $S_{tot}=2$ 
state for the pair it connects, and the
ground state of the entire chain is the spin fully polarized ferromagnetic
state. For $-\pi/4 < \theta < \pi/4$ the system is in the Haldane
gapped phase\cite{haldane}. For $-3\pi/4 < \theta < -\pi/4$ the chain is
spontaneously dimerized\cite{klumper}, 
while an extended gapless phase has been predicted in the
region $\pi/4 < \theta < \pi/2$\cite{itoi}. Except for the ferromagnetic 
phase, the other phases all have singlet ground states with  
unbroken spin-rotational-symmetry, 
and we therefore refer to bonds in this region as antiferromagnetic.

In the presence of {\em strong} randomness, 
the system may be studied using a hierarchical real space RG approach.
We search for the
bond in the system with the largest gap separating its ground state and 
lowest energy excited state, say the bond coupling spins 2 and 3, with 
coupling constants $J_2$ and $D_2$ (see Fig. 2a). 
If the ground state of this bond
is a singlet ($-3\pi/4 < \theta_2 < \arctan {1\over 3}$), then
spins 2 and 3 form an inert singlet in the low-energy
states of the system, and mediate effective couplings between their 
neighboring spins 1 and 4, which may be calculated
using second order perturbation theory\cite{rossnote}:
\begin{eqnarray}
\tilde{J}_{14}&=&{(2J_1-D_1)(2J_3-D_3)\over 3(J_2-3D_2)}
-{D_1D_3\over 9(J_2-D_2)};\\
\label{j14}
\tilde{D}_{14}&=&-{2D_1D_3\over 9(J_2-D_2)}.
\label{d14}
\end{eqnarray}
If the ground state of the bond is a triplet ($\arctan {1\over 3}
< \theta_2 < \pi/ 2$), 
spins 2 and 3 form an effective spin 2' with
$S_{2'}=1$,  
and its couplings to its neighbor spin 1 are 
\begin{eqnarray}
\tilde{J}_{12'}&=&(J_1-D_1)/2;\\
\label{j12}
\tilde{D}_{12'}&=&-D_1/2.
\label{d12}
\end{eqnarray}
Couplings to spin 4 have identical expressions. 
Should the ground state of the bond be a quintuplet, an effective
spin with $S=2$ forms, and the structure of the original spin-1
chain gets distorted.

Examination of Eqs. (2) and (\ref{d14})
shows that the generated bond may be {\em ferromagnetic},
even if all the bonds involved are {\em antiferromagnetic} and favor singlet
ground states. For concreteness, we consider the case where
bonds 2 and 3 are Heisenberg, i.e., $D_2=D_3=0$. In this case
the effective bond between spins 1 and 4 are: 
$\tilde{J}_{14}=2J_3(2J_1-D_1)/3J_2
$
and
$\tilde{D}_{14}=0$.
Therefore the generated bond is Heisenberg and {\em ferromagnetic} if
\begin{equation}
J_1-D_1/2 < 0.
\label{cond}
\end{equation}
Clearly, bond 1 can be AF and 
satisfy Eq. (\ref{cond}) if it lies in the shaded region of 
Fig. \ref{phase}.

We have verified the above results of perturbation theory, that it is possible
to get $ S_{tot} = 2 $ ground states when Eq. (\ref{cond}) is satisfied, by
performing exact diagonalization of four spin clusters. The results of one such
calculation is shown in Fig. \ref{exact}: as a function 
of $J_2$, the ground state
of the cluster changes from $S_{tot}=0$ (which evolves from a product of
singlets between spins $S_1$ \& $S_2$ and $S_3$ \& $S_4$ 
in the limit $J_2 = 0$)
to $S_{tot}=2$.

The perturbative RG described above
is reliable when the randomness is strong and the
distributions of the couplings are broad\cite{dm,fisher1}. 
We now consider the opposite limit of dilute randomness,
namely
a uniform AF spin-1 chain in the Haldane phase 
($-\pi/4 < \theta < \pi/4$, see Fig. \ref{phase}), with a finite Haldane gap
$\Delta$ and a small fraction of
impurity bonds that are much weaker than $\Delta$ (see Fig. 2b).
Here one must
identify the true low energy degrees of freedom\cite{hyman2}.
If the impurity bonds were taken away, the original chain would have been
chopped into decoupled segments; the low-energy degrees of freedom are
the two spin-1/2 at the two edges of each segment
with a coupling (that can be either
F or AF) decreasing 
exponentially with the length of the segment.
Putting back the impurity bonds, as long as they are weak compared to $\Delta$, does not 
alter the bulk structure of the segments; their primary effect is to couple
neighboring edge spins in {\em different} segments. 
Let us assume bond 1 coupling spins 1 and 2 is such an impurity bond, with
$J_1, D_1 \ll \Delta$. To calculate the coupling between
the two edge spin-1/2s, which we label 1' and 2', we project 
the original operators onto the subspace of states below
the Haldane gap, i.e., states of the effective edge spins\cite{sorensen}:
\begin{equation}
\tilde{H}_{1'2'}=PH_{12}P=J_1P{\bf S}_1\cdot{\bf S}_2P+D_1P
({\bf S}_1\cdot{\bf S}_2)^2P,
\end{equation}
where $P$ is the projection operator. Rotational symmetry as well as
properties of spin-1/2 guarantee that
$\tilde{H}_{1'2'}=\tilde{J}_{1'2'}{\bf S}_{1'}\cdot{\bf S}_{2'}+C_{1'2'}, 
$
where $C_{1'2'}$ is a constant. We also have $P{\bf S}_1\cdot{\bf S}_2P
=P{\bf S}_1P\cdot P{\bf S}_2P$, because spins 1 and 2 live in decoupled 
Hilbert spaces if bond 1 were not there, 
and the Wigner-Eckart theorem\cite{sakurai}
guarantees $P{\bf S}_iP=\alpha{\bf S}_i'$.
The constant $\alpha$ depends on bulk properties of the segments; 
for infinitely long segment with Heisenberg coupling, 
$\alpha\approx 1.0640$\cite{white}.
Similarly 
$P({\bf S}_1\cdot{\bf S}_2)^2P=a+b{\bf S}_{1'}\cdot{\bf S}_{2'},
$
and the important constant $b$ may be determined by calculating certain
matrix elements of $({\bf S}_1\cdot{\bf S}_2)^2$ in the subspace:
$b=2[\langle\uparrow_{1'}\uparrow_{2'}|({\bf S}_1\cdot{\bf S}_2)^2|
\uparrow_{1'}\uparrow_{2'}\rangle
-\langle\uparrow_{1'}\downarrow_{2'}|({\bf S}_1\cdot{\bf S}_2)^2|
\uparrow_{1'}\downarrow_{2'}\rangle].
$
Using the commutation relations of ${\bf S}$ and the fact\cite{white}
$\langle\uparrow_{1'}|S^z_1|\uparrow_{1'}\rangle=\alpha/2
$
etc, we obtain
$b=-\alpha^2/2$. Therefore
$\tilde{J}_{1'2'}=\alpha^2(J_1-D_1/2)$,
and again we find
$\tilde{J}_{1'2'}$ may be {\em ferromagnetic} even when the original bond 1 was AF, with the same condition as before, namely Eq. (\ref{cond}).
We have thus demonstrated the existence of such an instability to form
ferromagnetic couplings in both the high disorder and dilute disorder limits.

We now discuss the origin of this ferromagnetic instability,
and in particular, the significance of the 
special combination of $J$ and $D$ in Eq. (\ref{cond}). 
By introducing a new coupling constant
$K=J-D/2,
$
the RG equations (2) and (4) simplify significantly:
\begin{equation}
\tilde{K}_{14}={4K_1K_3\over 3(K_2-5D_2/2)};
\tilde{K}_{12'}=K_1/2.
\end{equation}
Combining these with equations (\ref{d14}) and (\ref{d12}), we find $K$ and 
$D$ {\em decouple} except through energy denominators, and one cannot generate
$K$ from $D$, or vice versa. This suggests that $K$ and $D$ represent 
couplings of operators with different symmetry properties.

To proceed further, we note that products of different components of the
spin operator (that appear in the Hamiltonian) may be organized to form
traceless irreducible spherical tensor operators\cite{sakurai}: 
$Y_{lm}({\bf S})$,
which is defined by replacing $\cos\theta$ by $S_z$, $\sin\theta\cos\phi$ by
$S_x$ and $\sin\theta\sin\phi$ by $S_y$ in the usual spherical harmonics
$Y_{lm}(\theta, \phi)$, and symmetrizing noncommuting 
components. $Y_{lm}({\bf S})=0$ for $l > 2S$.
A general way to write down the coupling between two spins that respect 
rotational symmetry (equivalent to 
$H_{ij}=\sum_{n=1}^{2S}J^{(n)}({\bf S}_i\cdot {\bf S}_j)^n$) is
\begin{equation}
H_{ij}=\sum_{l=0}^{2S}K^{(l)}\sum_{m=-l}^{l}(-1)^mY_{lm}({\bf S}_i)
Y_{l,-m}({\bf S}_j),
\end{equation}
where $K^l$ is the coupling constant of rank $l$ spherical tensors. It
is easy to verify that
\begin{eqnarray}
&&{\bf S}_1\cdot{\bf S}_2={4\pi\over 3}\sum_{m=-1}^1
(-1)^mY_{1m}({\bf S}_1)
Y_{1,-m}({\bf S}_2),\\
&&({\bf S}_1\cdot{\bf S}_2)^2=-
{2\pi\over 3}\sum_{m=-1}^1(-1)^mY_{1m}({\bf S}_1)Y_{1,-m}({\bf S}_2)\nonumber\\
&+&{8\pi\over 15}\sum_{m=-2}^2(-1)^mY_{2m}({\bf S}_1)Y_{2,-m}({\bf S}_2)
+{1\over 3}{\bf S}_1^2{\bf S}_2^2.
\end{eqnarray}
Therefore for the coupling of the form $J{\bf S}_1\cdot{\bf S}_2+
D({\bf S}_1\cdot{\bf S}_2)^2$, we have 
$K^{(1)}={2\pi\over 3}(2J-D)\propto K$, and $K^{(2)}={8\pi\over 15}D$.
Thus the $K$ variable is proportional to the coupling between rank 1 tensors
(vectors),
nothing but the Heisenberg coupling (in this tensor
representation); and Eq. 
(\ref{cond}) indicates the Heisenberg coupling is {\em ferromagnetic}.

The advantage of writing the Hamiltonian in terms of couplings of
irreducible spherical tensors instead of powers of ${\bf S}_1\cdot{\bf S}_2$
is that different symmetry properties of tensors with different ranks do 
not allow them to mix in 1st and 2nd order perturbation calculations, as we
have already seen; while $({\bf S}_1\cdot{\bf S}_2)^n$ in general includes
tensor couplings with all ranks up to $n$. For example, we know 
$Y_{lm}({\bf S})$ acting on a singlet creates an eigenstate 
of ${\bf S}^2_{tot}$
and $S^z_{tot}$ with $S_{tot}=l$ and $S^z_{tot}=m$, therefore in the second
order perturbation a coupling of rank $l$ between spins 1 and 4 is mediate 
through the channel of excited states with $S_{tot}=l$
of spins 2 and 3.
More remarkably, when projecting spin-1 couplings to couplings between
spin-1/2 edge spins, the fact that spin-1/2 object does not support 
tensors with ranks higher than 1 guarantees that the coupling must be 
proportional to $J-D/2$ of the original coupling, and the 
original rank 2 tensor 
coupling simply gets eliminated.

The above discussion also gives us insights into the origin of
the ferromagnetic instability we demonstrated. Even though a single bond
favors a singlet ground state, it may well contain ferromagnetic couplings
between tensors with certain rank (say, rank 1, or Heisenberg coupling).
In the absence of randomness no ferromagnetic instability is triggered by
such couplings. In the presence of randomness, however, energy scales in 
the system gets separated, and low- and high- energy subspaces of the
Hilbert space get perturbatively decoupled (which is the basis for perturbative
RG). In projecting to low-energy subspaces, certain AF couplings may get
suppressed for symmetry reasons (as we have seen), while the original 
subdominant ferromagnetic couplings may survive and become dominant. 
In the example we illustrated above, the AF rank 2 couplings are suppressed
in the low energy subspace, while the ferromagnetic rank 1 (Heisenberg)
coupling survives.
This is the origin of the ferromagnetic instability, which is
a purely quantum mechanical effect, and absent without randomness.

With the generation of ferromagnetic bonds {\em between} segments as well as
within segments, the decimated chain is very different from that obtained with
pure Heisenberg coupling by Hyman and Yang\cite{hyman2}. 
In fact, it becomes of the
universality class of the spin-1/2 chain with random AF and F couplings
studied by Westerberg {\em et al.}\cite{westerberg}. Using our formalism,
their RG scheme may be easily generalized to include higher spins and
non-Heisenberg couplings, as flows of couplings of different rank tensors
tend to 
decouple. Our results show that at low energies high rank couplings are
strongly suppressed compared to Heisenberg couplings, if the original couplings
are dominantly Heisenberg, which is a likely situation in nature\cite{bhatt}.
Therefore the active degrees of freedom in our case will be weakly coupled
large moments as in the case of Westerberg{\em et al.}, which lead to a
pure Curie susceptibility in the low temperature limit, because of a
perfect cancellation between growing moments $ \mu $ and reducing numbers $ N $
in the expression for the susceptibility $ \chi = N(T) \mu (T) ^2 /T $
as $T$ is lowered.

In summary, we have demonstrated the possibility of generating ferromagnetic
couplings in quantum spin chains with 
{\em random} rotationally invariant couplings,
starting from the antiferromagnetic sector of the phase diagram of the pure
chain.
Though we have discussed $S=1$ exclusively for concreteness, similar
effects would be expected for higher spins, though their phase diagrams
even for the pure case are not fully known. 
For integer spins, our arguments are valid in both large disorder and dilute
limits, and the former case applies to larger half-odd-integer spins as well.
However, as our treatment shows, the effects are purely quantum mechanical,
and dependent on either the spin gap or the region of validity of second order
perturbation theory, both of which are known to become smaller with increasing
$S$. Consequently, we expect the region of this anomalous behavior to decrease
with increasing $S$, and disappear in the classical limit $S\rightarrow\infty$.

This work was supported by NSF grant
DMR-9400362 (at Princeton), and a Sherman Fairchild 
fellowship (at Caltech). 
Part of this work was performed at the Aspen Center for Physics.

\begin{figure}
\caption{Illustration of spin decimation procedures. (a) Strong randomness
case. When the strongest bond (between spins 2 and 3) has a singlet 
ground state, spins 2 and 3 are decimated and an effective bond connecting
1 and 4 is generated; 
when the ground state is a triplet, 2 and 3 form an effective 
spin-1 object $S_{2'}$, which is coupled to its neighbors 1 and 4. 
(b) Dilute randomness case. When a uniform spin-1 chain in the 
Haldane phase is broken into finite segments coupled by weak impurity bonds
(dotted lines), the low energy degrees of freedom are half spins
living at the edges of each segments; there is weak coupling between 
neighboring half spins, both in the same segment (broken lines) and 
different segments (dotted lines).
}
\label{rg}
\end{figure}

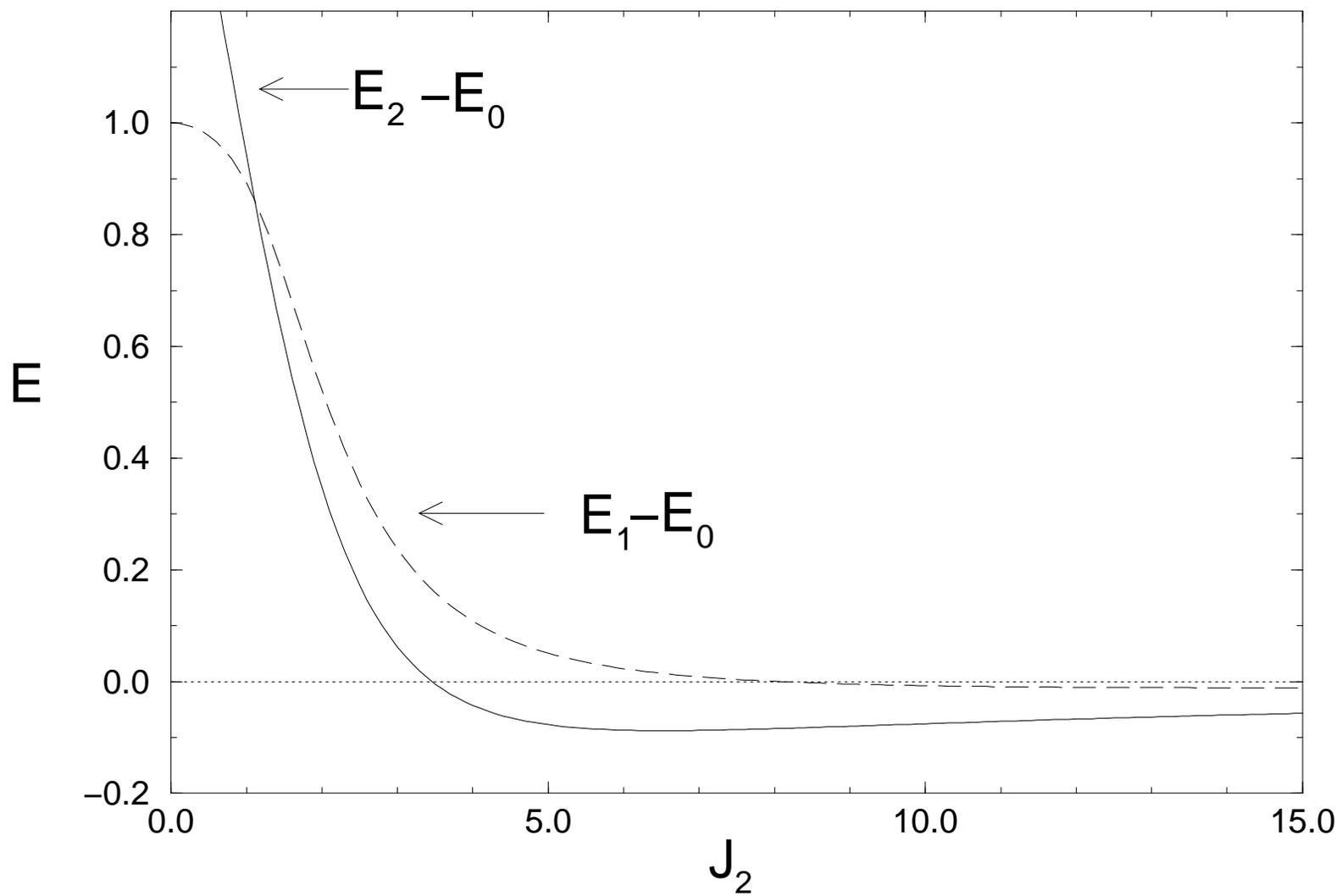
\begin{figure}
\caption{Ground state level crossing in a four-spin cluster for
$J_1=-1.0$, $D_1=-1.5$, $J_3=1.0$, $D_2=D_3=0$.
The plot shows energies of the lowest energy quintuplet ($S_{tot}=2$, 
solid line) and
triplet ($S_{tot}=1$,
broken line) states, measured from lowest energy singlet state.
For $J_2 > 3.46$, the ground state has total spin $S_{tot}=2$, 
despite the fact that all three bonds favor singlet ground states.
}
\label{exact}
\end{figure}

\begin{references}
\bibitem{note} In real systems the spin rotational symmetry may be broken
by crystal fields, and the symmetry may be reduced.
\end{references}
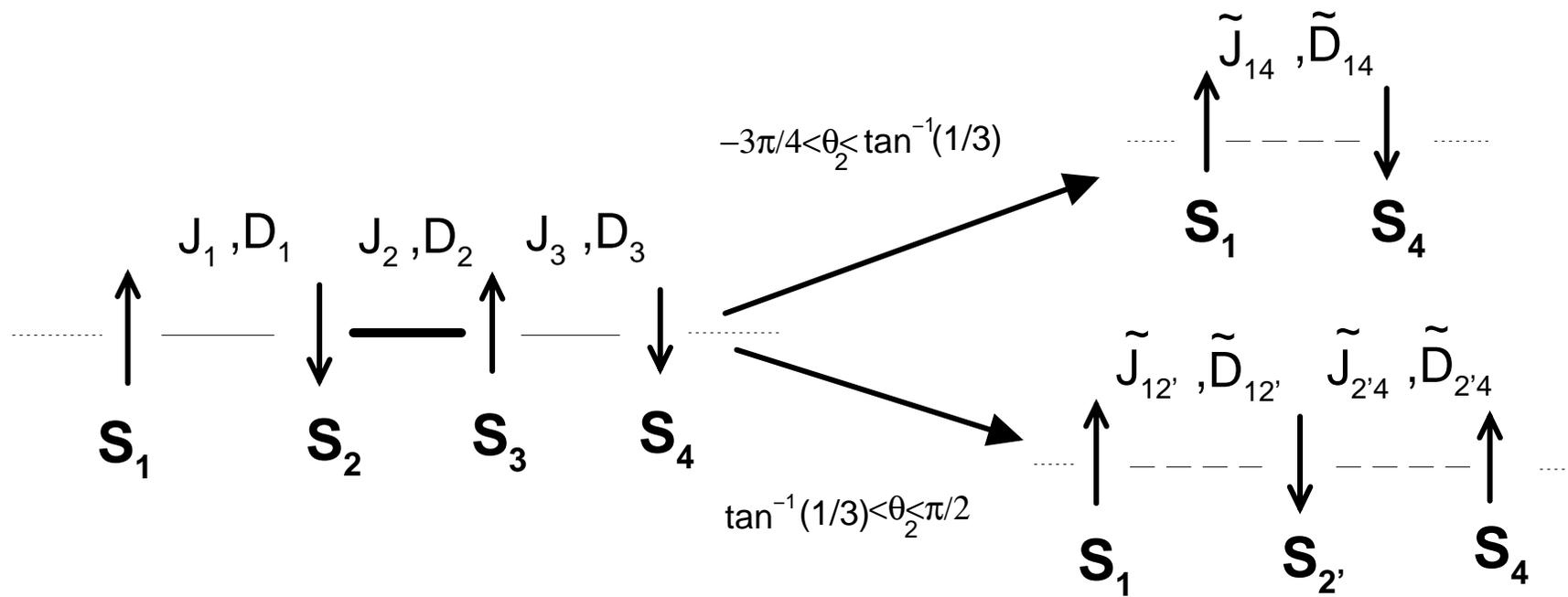
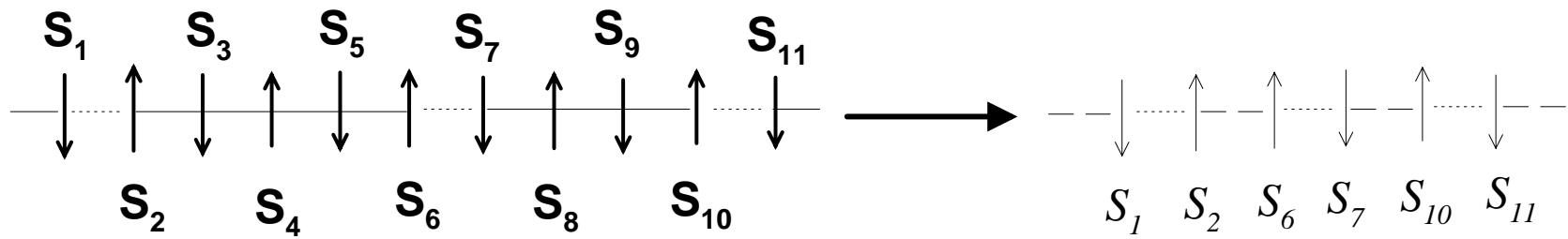
\end{document}